\documentclass[twocolumn,showpacs,preprintnumbers]{revtex4}
\usepackage{amssymb}
\usepackage{graphicx}
\usepackage{dcolumn}
\usepackage{bm}

\begin{document}

\title{Directional photoelectric current across the bilayer graphene junction}
\author{S.~E.~Shafranjuk}
\homepage{http://kyiv.phys.northwestern.edu}
\affiliation{Department of Physics and Astronomy, Northwestern University, Evanston, IL
60208}
\date{\today }

\begin{abstract}
A directional photon-assisted resonant chiral tunneling through a bilayer graphene barrier
is considered. An external electromagnetic field applied to the barrier switches the transparency $T$ in the longitudinal direction from its steady state value $T=0$ to the ideal $T=1$ at no energy costs. The
switch happens because the a.c. field affects the phase correlation between the
electrons and holes inside the graphene barrier changing the whole angular
dependence of the chiral tunneling (directional photoelectric effect). The
suggested phenomena can be implemented in relevant experiments and in
various sub-millimeter and far-infrared optical electronic devices.
\end{abstract}

\maketitle

\section{Introduction}
\begin{figure}[tbp]
\includegraphics{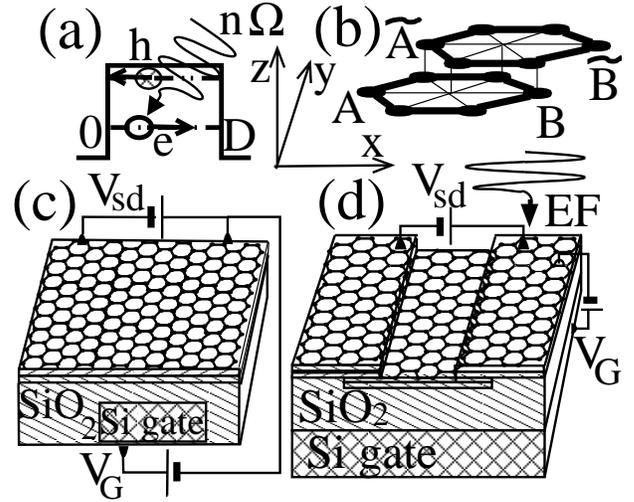}
\caption{(a) Potential barrier in the bilayer graphene controlled by the gate voltage $V_{\mathrm{G}}$ and exposed to the external electromagnetic field EF. The scattering states inside the barrier
originate from conversion of an electron ($e$) to a hole ($h$). (b) Two
coupled hexagonal lattices with non-equivalent carbon atomic sites A, B, 
\~{A}, and \~{B} in the bottom and top layers respectively. Two possible
setups (c) and (d) of the bilayer graphene junction. The external a.c.
field induces the directional photon-assisted resonant tunneling.}
\label{fig:Setup_b}
\end{figure}

Electromagnetic properties of the bilayer graphene\cite{Kats-C,Novos,McCann,Nilsson}
offer enormous opportunities for scientific research and various
nanoelectronic applications. They emerge in spectroscopy of bound and
scattering states, in the photon-assisted chiral tunneling and in direct
probing of strong correlation effects. Potential applications include
electromagnetic field (EF) spectral analyzers, receivers, detectors, and
sensors\cite{My-PRB}. The crystal lattice of the bilayer graphene\cite%
{Kats-C,Novos,McCann,Nilsson} consists of four equivalent sublattices of carbon
atoms while the charge carriers behave there as massive "chiral fermions"%
\cite{Kats-C,McCann,Nilsson}. The chiral fermions (CF) in bilayer graphene have a
finite mass $m_{e,h}$, like conventional electrons ($e$) and holes ($h$) in
metals and semiconductors\cite{Kats-C,McCann,Nilsson}. The chirality relates the
particles to certain sublattice and is responsible for various
unconventional d.c. electronic and magnetic properties of the bilayer
graphene\cite{Kats-C,Novos,McCann}. In contrast to an ordinary tunneling
through a conventional potential barrier, during the chiral tunneling (CT) an
incoming electron is converted into a hole moving inside the graphene
barrier in a reverse direction as indicated in Fig. \ref{fig:Setup_b}(a)
(Klein paradox\cite{Strange,Krekora}). This yields a finite transparency $%
T\neq 0$ for incident electrons with energies $E$ below the barrier $E<U_{0}$ ($U_{0}$ is the barrier height
energy) occurring\cite{Kats-C} at finite particle incidence angles $\phi \neq 0$. On the other hand, the steady state chiral tunneling is blocked ($T=0$) in the longitudinal direction $\phi =0$. The angle-dependent transparency makes the chiral tunneling being attractive for various nanoelectronic applications\cite{My-PRB,Torres}. The potential barrier in graphene can either be induced by the gate voltage $V_{\mathrm{G}}$ from a Si gate slab
or can be formed by three overlapping graphene sheets as shown in Figs. \ref%
{fig:Setup_b}(c,d). According to Ref. \cite{Kats-C}, the d.c. gate voltage $%
V_{\mathrm{G}}$ shifts the graphene barrier height, which controls the
chiral tunneling. That process implies the wavefunction phases of electrons and holes being
interconnected with each other in the graphene. The phase correlations during the chiral tunneling can
also be directly tuned by applying of an external a.c. field. Controlling of the electron wavefunction phase by an a.c. field had not been accomplished yet and is the subject of this paper. The electronic properties are described by a spinor wavefunction $\hat{\Psi}$, which components depend on the angle $\phi $ between the electron momentum $%
\mathbf{p}$ and the $x$-axis (see Fig.~\ref{fig:Setup_b}). Similar spinor
description had formerly been used for Dirac fermions\cite{Strange} and for
relativistic quasiparticles in single-layer graphene\cite{Kats-C,Novos2}. 

This paper is devoted to electromagnetic properties of a bilayer graphene
junction shown in Fig.~\ref{fig:Setup_b}. One may
expect that the differential tunneling conductance $\sigma (\phi
,V_{\rm sd}) $ of "clean" samples depends on the angle $\phi $ between the
electric current $\mathbf{j}$ and the $x$-direction (see Fig.~\ref%
{fig:Setup_b}). The whole shape of $\sigma (\phi )$ versus the
source-drain voltage $V_{\rm sd}$ is very sensitive to properties of the bilayer
graphene barrier. We begin with computing of the steady state $\sigma
(\phi ,V_{\rm sd})$ curves for a graphene barrier biased by $V_{\rm sd}$%
. The steady state results are then utilized for studying of the
a.c. properties. When an external electromagnetic field (EF) is applied, it
strongly affects the directional diagram of $\sigma (\phi ,V_{\rm sd})$. In
particular we will see that the external electromagnetic field induces a
finite conductance in the straightforward direction ($\phi =0$), which had been
blocked in the steady state. That happens because the electromagnetic field
affects the electron-hole phase correlations inside the graphene barrier
directly. In the steady state, when the a.c. field is off, the electric
current is fully suppressed at $V_{\rm sd}<U_{0}$ (for typical gate voltage $V_{\rm G}=1 $ V and the SiO$_{2}$ thickness $d=300$ nm one finds\cite{Novos2} $U_{0}=2 $ meV).

\section{Photon-assisted chiral tunneling}
Here we examine influence of an electromagnetic field to chiral
tunneling and discuss the intrinsic noise. For studying of the
non-stationary electric current across the bilayer graphene junction we
implement methods\cite{Yeyati,Keldysh,Datta}. The graphene bilayer is
modelled as two coupled gexagonal lattices consisting of four non-equivalent
sites A, B and \~{A}, \~{B} in the bottom and top layers respectively [see
Fig.~\ref{fig:Setup_b}(a)]. The chiral fermion Hamiltonian operates in space
of the two-component wave functions $\hat{\Psi}$. When the junction is
exposed to an external electromagnetic field, the main part of the Hamiltonian is 
\begin{equation}
\hat{H}=-\hbar ^{2}\left( \pi _{-}^{2}\hat{\sigma}_{+}+\pi _{+}^{2}\hat{%
\sigma}_{-}\right) /2m+U(x),  \label{H_0}
\end{equation}%
where $\pi _{\pm }=\left( k-eA_{x}\left( t\right) /\hbar \right) \pm i\left(
q-eA_{y}\left( t\right) /\hbar \right) $, $\hat{\sigma}_{\pm }=\hat{\sigma}%
_{x}\pm \hat{\sigma}_{y}$, $\hat{\sigma}_{i}$ are the Pauli matrices, $%
i=\{x,y,z\}$, the effective mass $m$ is expressed via coupling strength $%
\gamma _{\tilde{A}B}$ between $\tilde{A}$ and $B$ as $m=\gamma _{\tilde{A}%
B}/2v^{2}=0.054$ $m_{e}$, where $v=(\sqrt{3}/2)a\gamma _{\mathrm{AB}}$, $%
a=0.246$ nm is the lattice constant, $\gamma _{\mathrm{AB}}\approx 0.4$ eV, $%
A_{x,y}\left( t\right) $ are corresponding components of the time-dependent
vector potential $\mathbf{A}\left( t\right) $, $U(x)$ is the graphene
barrier potential controlled by the gate voltage $V_{\mathrm{G}}$. Eq. (\ref%
{H_0}) describes interlayer coupling via a dimer state formed by pairs of
carbon A\~{B} atoms located in the bottom and top layers respectively as
shown in Fig.~\ref{fig:Setup_b}(b). A weak direct A\~{B} coupling and a
small interaction due to the bottom and top layer asymmetry (which opens a
minigap in the electron spectrum\cite{McCann}) are both hereafter neglected.

For graphene junctions having finite dimensions, the motion of chiral
fermions is quantized. The quantization imposes additional constrains on the
directional tunneling diagram. Permitted values of the angle $\tilde{\phi}%
_{n}$ inside the graphene barrier are obtained from boundary conditions
along the $y$-direction, so the $y$-component of the electron momentum $%
\mathbf{p}=(\hbar k,\hbar q)$ is quantized as $\tilde{q}_{n}=n\pi /W$ (where 
$W$ is the barrier width), which gives $\tilde{\phi}_{n}=\arctan \left[ n\pi
/(k_{\varepsilon }^{\prime }W)\right] $ where $k_{\varepsilon }^{\prime }=%
\sqrt{2m/\hbar ^{2}}\sqrt{\left\vert \varepsilon -U_{0}\right\vert
-\left\vert \varepsilon \right\vert \left( 1-\cos 2\phi \right) /2}$. The
last formula also means that $\tilde{q}_{n}$ depends on the electron energy
variable $\varepsilon $. The electric current density $j=I(V_{\rm sd})/W$ ($I$ is the electric current, $V_{\rm sd}$ is the bias voltage, and $W$ is the graphene stripe width) between the electrodes 1 and 3
is computed as $j=2\pi e\int d\varepsilon \chi _{\varepsilon }\left[
G_{3}^{K}\left( \varepsilon \right) -G_{1}^{K}\left( \varepsilon \right) %
\right] $ where we introduced the factor $\chi _{\varepsilon }$. If 1 and 3
electrodes are made of a monolayer graphene or are metallic, then $\chi
_{\varepsilon }=v_{F}N\left( 0\right) $ where $v_{F}$ and $N\left( 0\right) $
are corresponding Fermi velocity and the electron density of states at the
Fermi level. However if the 1,3 electrodes are made of the bilayer graphene
itself, which case we inspect in details below, then $\chi _{\varepsilon
}=v_{\varepsilon }N\left( \varepsilon \right) $ where $v_{\varepsilon
}=\hbar \left\vert k\right\vert /m=\sqrt{2\left\vert \varepsilon \right\vert
/m}$ and $N\left( \varepsilon \right) =\sum_{k}\theta \left( \varepsilon
-E_{k}\right) \cdot m/\left( \pi \hbar ^{2}\right) $ are the energy
dependent velocity and the two-dimensional electron density of states in the
bilayer graphene, $E_{k}$ is the $k$-th electron energy level in the
graphene barrier stripe, $G_{r}^{K}\left( \varepsilon \right)
=-i\sum_{p}\left\vert t_{p}\right\vert ^{2}e^{iqy}e^{ikD}\left(
2n_{p}-1\right) \delta \left( \varepsilon -\varepsilon _{p}+\delta
_{r,3}eV_{\rm sd}\right) $ is the Keldysh Green function\cite{Keldysh}, $r$ is
the electrode index, $\delta _{r,3}$ is the Kronecker symbol, $n_{p}$ is the
distribution function of electrons with momentum $\mathbf{p}$. A
straightforward calculation using methods of Refs. \cite%
{Yeyati,Keldysh,Datta} gives 
\begin{eqnarray}
j &=&(\pi /2)e\int d\varepsilon \chi _{\varepsilon }\sum_{p}\left\vert
t_{p}\right\vert ^{2}[(2n_{p}-1)  \nonumber \\
&&\cdot \delta (\varepsilon -\varepsilon _{p}+eV)-(2n_{p}-1)\cdot \delta
(\varepsilon -\varepsilon _{p})]  \nonumber \\
&=&\pi e\int d\varepsilon \chi _{\varepsilon }\left\vert t_{\varepsilon
}\right\vert ^{2}(n_{\varepsilon -eV}-n_{\varepsilon }).  \label{current}
\end{eqnarray}%
Taking for simplicity $N\left( \varepsilon \right) = m/\left( \pi \hbar ^{2}\right) $ from Eq. (\ref{current}) one finds the zero-temperature steady-state conductance as
\begin{eqnarray}
G_{0}=\frac{e^{2}}{\hbar ^{2}}\overline{T}W\sqrt{2meV_{\rm sd}} =\frac{2e^{2}}{h} \overline{T} N_{\rm ch} \left( V_{\rm sd}\right) 
\label{G_0}
\end{eqnarray} 
where $\overline{T}=\left\vert t_{eV_{\rm sd}}\right\vert ^{2}$ is  the graphene barrier transparency. In Eq.~(\ref{G_0}) we introduced the voltage-dependent dimensionless number of conducting channels $N_{\rm ch} \left( V_{\rm sd}\right) =\pi W \sqrt{2meV_{\rm sd}} $. The dependence $N_{\rm ch}$ versus $V_{\rm sd}$ stems from the energy dependence of the electron velocity in the bilayer graphene $v_{\varepsilon }$.  Eq.~(\ref{G_0}) coincides with well known Landauer formula with the number of conducting channels $N_{\rm ch}$. The calculation results will be convenient to normalize to an auxiliary conductivity defined as $\tilde{\sigma}_{0}=W^{-1} \cdot G_{0}(V_{\rm sd}=U_0/e)=(2e^{2}/h) \pi \sqrt{2mU_0}$ (where we used $\overline{T} \simeq 1$ at $V_{\rm sd}=U_0/e$, $U_0$ being the graphene barrier height). The transmission amplitude $t_{\varepsilon }$ across the voltage biased junction is obtained within a simple model which represents the chiral fermion wavefunctions via Airy functions. The Hamiltonian (\ref{H_0}) yields a gapless semiconductor with massive chiral electrons and holes having a finite mass $m$. Let us consider tunneling of those fermions with the energy $E$ incident on the barrier under the angle $\phi $. Since the potential barrier is formed in the longitudinal direction, the $y$-component $\hbar q$ of the momentum $\mathbf{p}$ is conserved while the $x$-component $\hbar k$ is not. The trial chiral fermion wavefunction takes a piece-wise form\cite{Kats-C}. The chirality has no significance for particles propagating above the barrier $E>U_{0}$. An analytical steady state solution\cite{Kats-C} is obtained at $V_{\rm sd}=0$ for a rectangular barrier expressing the electron and hole wavefunctions via combinations of plane waves. Matching the continuous boundary conditions one finds\cite{Kats-C,Novos,McCann,Nilsson} the tunneling amplitude $t_{\rm 2GW}$ for a normal electron incidence ($\phi=0$) as
\begin{eqnarray}
t_{\rm 2GW}=-\frac{2\text{$k$}(\text{$k^{\prime }$}-\text{$k$})se^{2i(D\text{$k^{\prime }$}+2%
\text{$\varphi $})}}{e^{2iD\text{$k^{\prime }$}}(\text{$k$}-\text{$k^{\prime }$})^{2}\text{$%
s^{\prime }$}-(\text{$k$}+\text{$k^{\prime }$})^{2}\text{$s^{\prime }$}}
\label{2GW}
\end{eqnarray}
where the electron wave vector in the electrode is $k=\sqrt{2m|E|} /\hbar $ and inside the barrier is $k^{\prime }=$ $\sqrt{2m(E-U_{0})}/\hbar $, $\varphi $ is the phase drop across the graphene barrier, $s^{\prime }={\rm sign}{(U_0-E)}$. For a classic rectangular barrier one instead obtains
\begin{eqnarray}
t_{cl}=\frac{k\text{$k^{\prime }$}e^{-iDk}e^{i\varphi }}{k\text{$k^{\prime }$}\cos (D\text{$%
k^{\prime }$})-i\left( k^{2}+\text{$k^{\prime }$}^{2}\right) \sin (D\text{$k^{\prime }$})/2}.
\label{classic_barr}
\end{eqnarray}
Although Eqs.~(\ref{2GW}), (\ref{classic_barr}) are instructive, the experimentally measured characteristics are relevant rather to a finite bias voltage ($V_{\rm sd} \neq 0$) across the graphene barrier and finite incidence angles $\phi \neq 0$. The electric field $\mathcal{E}$ in the latter case penetrates inside the bilayer graphene barrier and electrodes, forcing the charge carriers to accelerate. Simplest electron and hole
wavefunctions in that case are represented via the Airy functions\cite{Korn} rather than via plane waves. The CF wavefunction $\hat{\Psi}\left( x\right) $ is obtained from the Dirac equation $\hat{H}\hat{\Psi}=E\hat{\Psi}$ where $E$
is the electron energy. For calculations one uses the tilted barrier potential $U\left( x\right) =-\mathcal{E}x\left[ \theta \left( -x\right) +\theta \left( x-D\right) \right] $ $+\left[ U_{0}-\mathcal{E}x\right] \theta \left( x\right) \theta \left( D-x\right) $ where $\mathcal{E}=V_{\rm sd}/D$ is the electric field, which penetrates into the graphene barrier. Then components of the fermion momentum $\mathbf{p}=(\hbar k,\hbar q)$ are written as $\hbar q=\sqrt{2m\left\vert E\right\vert }\sin {\phi }$ and $\hbar k\left( x\right) =i \sqrt{2m(U\left( x\right) -E)}\cos \phi \left( x\right) $, $\phi \left(
x\right) =\arcsin {[}\left( {q/}k\left( x\right) \right) {\sin {\phi }]}$ where $D$ is the barrier thickness, $\phi $ is the electron incidence angle in the electrode 1. The corresponding trial wavefunction is 
\begin{eqnarray}
\hat{\Psi} &=&\hat{\Psi}_{1}\theta (-x)+\hat{\Psi}_{2}\theta (D-x)+\hat{\Psi}%
_{3}\theta (x-D)  \nonumber \\
\hat{\Psi}_{1} &=&e^{iqy}[\lambda \mathrm{Bi}(\zeta _{k,x})+b_{1}\tilde{%
\lambda}\mathrm{Bi}(\zeta _{k,x})+c_{1}\lambda ^{\dag }\mathrm{Ai}(\zeta
_{ik,x})]  \nonumber \\
\hat{\Psi}_{2} &=&e^{iqy}[a_{2}\mathrm{Ai}(\zeta _{k^{\prime },x})\mu +b_{2}%
\mathrm{Bi}(\zeta _{k^{\prime },x})\tilde{\mu}  \label{trial_f} \\
&&+d_{2}\mathrm{Bi}(\zeta _{ik^{\prime },x})\mu ^{\dag }+c_{2}\mathrm{Ai}%
(\zeta _{ik^{\prime },x})\mu ^{\ddag }]  \nonumber \\
\hat{\Psi}_{3} &=&e^{iqy}[a_{3}\mathrm{Ai}\left( \zeta _{k,x}\right) \nu
+d_{3}\mathrm{Bi}\left( \zeta _{ik,x}\right) \tilde{\nu}]  \nonumber
\end{eqnarray}%
where $\zeta _{k,x}=-\left( k^{2}+\mathcal{E}x\right) /\left( -\mathcal{E}%
\right) ^{2/3}$, $k=\sqrt{2m|E|}\cos \phi /\hbar $ is the electron wave vector in the electrode, $k^{\prime }=$ $\sqrt{2m(E-U_{0})}\cos \phi ^{\prime }/\hbar $ is the electron wave vector inside the graphene barrier, $\phi ^{\prime }=\arcsin \left((q/k^{\prime })\sin {\phi }\right) $, $s_{1}=-1$ , $s_{2}=\mathrm{sign}%
\left( U_{0}-E\right) $, $s_{3}=\mathrm{sign}\left( -V_{\rm sd}-E\right) $, $%
h^{\prime }=\sqrt{1+\sin ^{2}\phi ^{\prime }}-\sin \phi ^{\prime }$, $%
\lambda =(\left\vert \uparrow \right\rangle +s_{1}e^{2i\phi }\left\vert
\downarrow \right\rangle )$, $\tilde{\lambda}=(\left\vert \uparrow
\right\rangle +s_{1}e^{-2i\phi }\left\vert \downarrow \right\rangle )$, $%
\lambda ^{\dag }=(\left\vert \uparrow \right\rangle +s_{1}h_{1}\left\vert
\downarrow \right\rangle )$, $\nu =\left( \left\vert \uparrow \right\rangle
+s_{3}e^{2i\phi }\left\vert \downarrow \right\rangle \right) $, $\tilde{\nu}%
=\left( \left\vert \uparrow \right\rangle -s_{3}/h_{3}\left\vert \downarrow
\right\rangle \right) $, $\mu =(\left\vert \uparrow \right\rangle
+s_{2}e^{2i\phi ^{\prime }}\left\vert \downarrow \right\rangle )$, $\tilde{%
\mu}=(\left\vert \uparrow \right\rangle +s_{2}e^{-2i\phi ^{\prime
}}\left\vert \downarrow \right\rangle )$, $\mu ^{\dag }=(\left\vert \uparrow
\right\rangle -s_{2}/h_{2}\left\vert \downarrow \right\rangle )$, $\mu
^{\ddag }=(\left\vert \uparrow \right\rangle -s_{2}h_{2}\left\vert
\downarrow \right\rangle )$. In the above equations we introduced auxiliary
matrices $\left\vert \uparrow \right\rangle ^{T}=(%
\begin{array}{cc}
1 & 0%
\end{array}%
)$ and $\left\vert \downarrow \right\rangle ^{T}=(%
\begin{array}{cc}
0 & 1%
\end{array}%
)$ (where $T$ means transpose). The chiral tunneling is pronounced at finite incidence angles $\phi \neq 0$ and at energies $E<U_{0}$ below the barrier. 
\begin{figure}[tbp]
\includegraphics{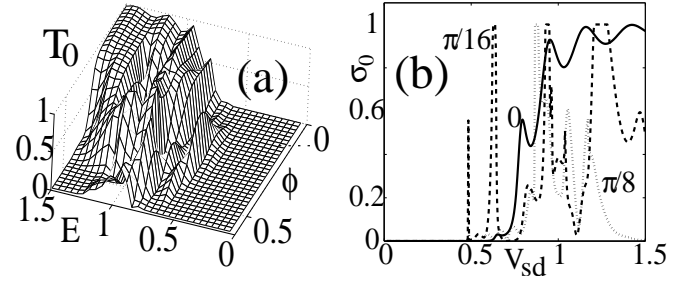}
\caption{(a)~The steady state tunneling transparency $T_0$ versus the electron
energy $E$ (in units of the graphene barrier height $U_{0}$) and the
azimuthal angle $\protect\phi $ (in radians). (b)~The corresponding steady
state differential conductance $\protect\sigma _{0}$ [in units of $\tilde{\sigma}_{0}=(2e^{2}/h) \pi \sqrt{2mU_0}$] versus the source-drain bias voltage $V_{\rm sd}$ (in units
of $U_{0}/e$) for three angles of incidence $\protect\phi $. The sharp peaks
at $V_{\mathrm{sd}}<U_{0}/e$ when $\protect\phi \neq 0$ originate from the
electron-hole interference inside the barrier.}
\label{fig:GQW1}
\end{figure}
The  steady state tunneling probability $T_{0}$ of a normally incident chiral particle vanishes below the barrier ($E<U_{0}$) while is finite above the barrier (when $E\geq U_{0}$). In Fig.~\ref{fig:GQW1}(a) we plot $T_0$ versus the energy $E$ of an electron incident to the barrier under the angle $\phi$. In Fig.~\ref{fig:GQW1}(b) we show the steady state tunneling differential conductance $\sigma_0 (V_{\rm sd})$ for different incidence angles $\phi$.  Both the plots in Figs.~\ref{fig:GQW1}(a,b) are related to $U_{0}=2$ meV, which corresponds to the surface charge density $n=10^{11}$ cm$^{-2}$ induced by the gate voltage $%
V_{\rm G}=1$ V across the SiO$_{2}$ substrate with thickness $d=300$ nm [see
Figs.~\ref{fig:Setup_b}(c,d)].

\begin{figure}[tbp]
\includegraphics{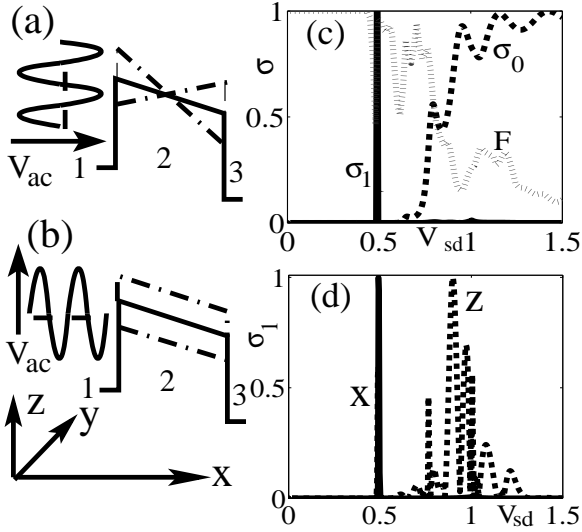}
\caption{Limit of a low a.c. field amplitude. Photoelectric effect in the bilayer graphene junction
induced by an external electromagnetic field. Modulation of the graphene
barrier height by the a.c field polarized along $\hat{x}$ (a) and $\hat{z}$
(b) axes. (c)~Corresponding steady state $\protect\sigma _{0}$ and the photon-assisted chiral tunneling
differential $\protect\sigma _{1}=\overline{\sigma}^t - \sigma_0$ conductances [same units as in Fig.~%
\protect\ref{fig:GQW1}(b)] in the longitudinal direction $\protect\phi =0$. Curve F shows the Fano factor $F$,
which characterizes the Poisson noise. (c) The d.c. conductance for $%
\hat{x}$- and $\hat{z}$- field polarizations. One may notice that the a.c.
field induces sharp resonant peaks in the photon-assisted chiral tunneling conductance $\protect%
\sigma _{1}$.}
\label{fig:GQW2}
\end{figure}

\section{Directional photo-electric current}
\begin{figure}[tbp]
\includegraphics{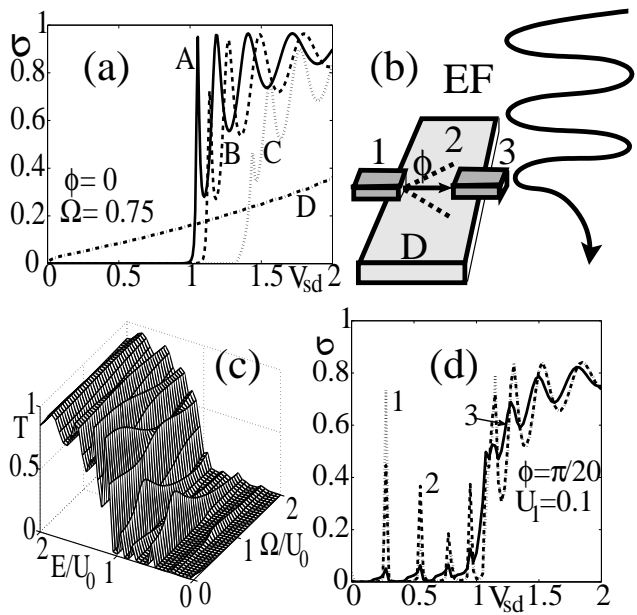}
\caption{The time averaged differential conductance $\overline{\sigma (t)}^{t}$ (in units of $\tilde{\sigma}_{0}=(2e^{2}/h) \pi \sqrt{2mU_0}$) of a bilayer graphene junction exposed to an external electromagnetic field
which modulates the barrier height $U(t)=U_{0}+U_{1}\cos {\Omega t}$. In
Fig.~4(c) one may notice a remarkably strong DPE at $\Omega /U_{0}\simeq 1$. This corresponds to curve 1 in Fig. 4(d) where the peak spacing $\Delta_k $ is determined by the graphene barrier length $D$.}
\label{fig:E_vs_W}
\end{figure}

The steady state characteristics of the d.c. biased graphene junction described above allow studying of the the external a.c. field influence to the graphene junction. We find that a most spectacular phenomenon occurs when
the a.c. gate voltage $V_{\mathrm{G}}\left( t\right) $ modulates the height $U$ of graphene barrier $U\rightarrow U_{0}+U_{1}\cos \Omega t$ where $\Omega $ is the a.c. field frequency. Then the $\hat{x}$-component of the electron momentum $\hbar k_{\mathrm{B}}=\sqrt{2m\left( U_{0}-E\right) }\cos \phi ^{\prime }$ inside the barrier becomes time-dependent $k_{\mathrm{B}}\rightarrow k_{\mathrm{B}}+\kappa \left( t\right) $ which at $\kappa \left(
t\right) <<k_{\mathrm{B}}$ gives $\kappa \left( t\right) =\left( U_{1}/2k_{\mathrm{B}}\right) \cos \Omega t+\left( U_{1}^{2}/8k_{\mathrm{B}}^{3}\right) \cos ^{2}\Omega t+O\left( U_{1}\right) $. We emphasize that a mere factorization\cite{My-PRB} of the electron wave function $\hat \Psi (x,t)$ like $\hat \Psi (x,t) \rightarrow \hat \Psi (x) \sum_k J_k \left(\alpha \right)\exp{(i n \Omega t)} $ [where $\alpha = e U_1/(\hbar \Omega )]$ is not working here since it does not properly incorporate the non-stationary behavior of $\hat \Psi (x,t)$. The puzzle comes from a non-analytical dependence of $\hat \Psi (x,t)$ on $\kappa \left( t\right)$. Therefore one should obtain a valid $\hat \Psi (x,t)$ from corresponding non-stationary boundary conditions at the electrode/barrier interfaces. This gives a complex non-stationary and non-linear behavior of $\hat \Psi (x,t)$, from which one computes the observable characteristics of interest. The a.c. field induced time dependence $\kappa \left( t\right) $ yields two spectacular consequences. First, the a.c. field splits the sharp resonance in the energy-dependent transmission probability $T(E)$ at $E=E_{0}$ as $E_{0}\rightarrow E_{0}\pm n\Omega $ where $n$ is the number of photons absorbed during the chiral tunneling process. And second, the a.c. field strongly affects the angular dependence of the chiral tunneling since it renormalizes the angle $\phi ^{\prime }$ between $q$ and $k_{\mathrm{B}}\left( t\right) $ inside the barrier as 
\begin{equation}
\phi ^{\prime }=\arcsin {[}\left( \overline{{q\cdot }k_{\mathrm{B%
}}^{-1}\left( t\right) }^{t}\right) {\sin {\phi }]}.
\label{phi}
\end{equation}
In order to compute the time-dependent electric current one solves the non-stationary boundary conditions. In this way one finds the transmission coefficient $t_{E}(t)$. Analytical expressions for $t_{E}(t)$ are obtained in a simplest case $U_{0}=0$ (no graphene barrier when the a.c. field is off). After the a.c. field is on, it induces an oscillating potential barrier $U\left( t\right) =U_{1}\cos \Omega t$ via an a.c. gate voltage $V_{\mathrm{G}}\left( t\right) =V_{\mathrm{G}}^{\left( 0\right) }\cos \Omega t$. Assuming a normal incidence ($\phi =0$) and setting $U_0=0$, $k_{2}=k_{1}+\kappa _{1}$ where  $\kappa _{1}$ is time-dependent one gets
\begin{eqnarray}
t_{E<U_{0}}\left( \phi =0\right) = \nonumber \\
\frac{4ik_{1}k_{+} \left( \cosh D\kappa _{-} +\sinh D\kappa _{-}\right) }{4ik_{1}k_{+}
\cosh Dk_{+} +2k_{1}^{2}\sinh Dk_{+}-2k_{+}^{2}\sinh Dk_{+}}
\label{trE}
\end{eqnarray}
where $\kappa _{-}=\kappa _{1}-ik_{1}$, $k_{+}=k_{1}+\kappa
_{1}$. Eq. (\ref{trE}) corresponds to a setup where the
graphene barrier is induced purely by the a.c. gate voltage. 
In the limit of small external a.c. field ($\kappa _{1}<<k_{1}$) from Eq. (\ref{trE}) one obtains 
\begin{eqnarray}
t_{E<U_{0}}\left( \phi =0\right) =2\mathcal{K}_{1}+2(D\text{$k_{1}$}%
+i)\left( 1-e^{2D\text{$k_{1}$}}\right)  \nonumber \\
\cdot \mathcal{K}_{2}\varkappa +(D^{2}e^{4D\text{$k_{1}$}}\text{$k_{1}$}%
^{2}+D^{2}\text{$k_{1}$}^{2}-12iDe^{2D\text{$k_{1}$}}\text{$k_{1}$} \nonumber \\
-6D^{2}e^{2D\text{$k_{1}$}}\text{$k_{1}$}^{2}+2iDe^{4D\text{$k_{1}$}}\text{%
$k_{1}$}+2iD\text{$k_{1}$} \nonumber \\
+4e^{2D\text{$k_{1}$}}-(2+i)-(2-i)e^{4D\text{$k_{1}$}})\mathcal{K}%
_{3}\varkappa ^{2}+O\left( \text{$\kappa _{1}$}^{3}\right) 
\label{t_En}
\end{eqnarray}%
where $\varkappa =\kappa _{1}(t)/k_{1}$ and we introduced the auxiliary
function $\mathcal{K}_{p}=e^{(1-i)D\text{$k_{1}$}}/\left( 1+e^{2D\text{$k_{1}
$}}\right) ^{p}$. The transmission resonances correspond to vanishing
denominator $\left( 1+e^{2D\text{$k_{1}$}}\right) ^{p}=0$, $p=\overline{%
1\dots 3}$. The Fourier transform of the above equation shows that the a.c.
field splits the $k-$th chiral tunneling resonance as $E_{k}\rightarrow
E_{k}\pm n\hbar \Omega $ where $n$ is the number of photons absorbed
(emitted) during the tunneling. One can see that the external field not only splits the resonances, but also strongly affects angular dependence of the chiral tunneling. That happens because the a.c. field causes no influence to the $\hat{y}$-component of the electron momentum $q$ since the graphene barrier
is effectively one-dimensional. The time dependence $\kappa _{1}\left(
t\right) $ takes also place when the a.c. field modulates the graphene
barrier width as $D\rightarrow D_{0}+D_{1}\cos \Omega t$. Splitting of the
chiral tunneling resonances, and the angular redistribution of the electric
current under the a.c. field influence is better pronounced for a finite barrier height $U_{0}\neq 0
$ and $U=U_{0}+U_{1}\cos \Omega t$. From Eq. (\ref{phi}) one can see that $\phi^{\prime }=0$ if $\phi=0$. However if $\phi \neq 0$, one may observe a spectacular phenomena. In this case an
external a.c. field induces a finite electric current for an almost normal
incidence $\phi \approx 0$, which was inhibited when the field was off. When $\phi \approx 0$, the a.c.
field actually causes additional photon-assisted chiral tunneling resonances to engage. The directional photoelectric effect (DPE) may be realized in two scenarios. \textit{One scenario} assumes that an electron beam having a finite angular width $\delta \phi \neq 0$ enters the graphene barrier normally.  A visible DPE can be achieved in the setup shown in Fig.~\ref{fig:E_vs_W}(b) where the attached electrodes 1,3 are made of one-dimensional conducting wires. If the wire is much narrower than the width of graphene stripe ($W_{\rm w} << W$), one may consider the electric current as a result of one-dimensional propagation of  of electron along the trajectories under influence of the bias voltage. Such method formerly had intensively been used in numerous works devoted to point contact junctions\cite{Kulik,Yanson}. If the electric current is sufficiently weak, the electrons coming from the wire into the graphene stripe introduce a negligible disturbance into the electron spectrum inside graphene. The translational invariance inside graphene is well preserved\cite{Wildoer}. Authors of Ref.~\onlinecite{Wildoer} used the STM tip for imaging of the electron wavefunction in carbon nanotube which showed a periodic pattern.  The electrode 1 emits electrons under a small but finite angle $\phi $ ($\phi <<\pi $, $\phi \neq 0$) which trajectories are focused/defocused by the external electromagnetic field as indicated in Fig.~\ref{fig:E_vs_W}(b). The frequency dependence of the transparency is governed by the directional photoelectric effect. A significant directional photo-electric effect emerges even for a relatively long wavelength $\lambda \simeq 1$~mm - $0.01$~$\mu $m (which corresponds to the THz domain) if the condition $\left\vert E-U_{0}\mp \Omega \right\vert <<\left\vert E\right\vert $ is met. The deviation angle $\phi ^{\prime }=\arcsin {[\sqrt{\left\vert E\right\vert /\left\vert E-U_{0}\mp \Omega \right\vert }\sin {\phi }]}$ inside the graphene barrier considerably increases giving $\phi ^{\prime }>>\phi $. This means that an ideal transparency taking place in the steady state at $\phi \neq 0$ is redistributed over the angle $\phi^{\prime }$ after the a.c. field is applied. The transparency peaks are actually shifted from finite angles $\phi \neq 0$ to the normal incidence angle $\phi = 0$.  \textit{Another scenario} involves an incident single electron which enters the graphene barrier strictly normally ($\phi =0$) under influence of a high frequency THz wave. In this scenario an electron absorbs a THz photon having the finite energy $E_{\Omega }$ and momentum $q$ along the y-axis. Then the electron deviation angle $\delta \phi $ just before entering the barrier is small, $\delta \phi << \pi $. For instance taking $\nu =30$ THz (which corresponds to the photon energy $E_{\Omega }=125\times 10^{-3}\mathrm{eV}$) one gets $\delta \phi \approx q/k=2\times 10^{-3}$. The photoelectric effect is well pronounced for an electron with energy $E_{e}\simeq 2\cdot 10^{-3}$ $\mathrm{eV}$ after it gets inside the graphene barrier. There if $\left\vert E-U_{0}\mp \Omega \right\vert <<\left\vert E\right\vert $ the deviation angle $\phi ^{\prime }=\arcsin {[\sqrt{\left\vert E\right\vert /\left\vert E-U_{0}\mp \Omega \right\vert }\sin {\phi }]}$ increases considerably, since the photon energy is pretty high, $E_{\Omega }/E_{e}\simeq 50$, $E_{\Omega
}=0.1$ \textrm{eV}. Practically this means that one must set $\hbar \Omega \simeq U_{0}$ to get a strong photoelectric effect. In the above example the last condition also supposes that one should use $U_{0}\approx E_{\Omega }=125$ meV. Below we consider two most important field polarizations along the $\hat{x}$ and $\hat{z}$ axes as shown in Figs.~\ref{fig:GQW2}(a,b). The barrier transparency $T(E,\phi )$ is affected by the a.c.
field directly in either case. In particular, the barrier shape is modulated by the a.c. field polarized along the $x$-direction as sketched in Fig.~\ref{fig:GQW2}(a), since $\mathcal{E}\rightarrow \mathcal{E}_{0}+\mathcal{E}_{1}\cos \left( \Omega t\right) $. On other side, if one applies an a.c. field polarized as $\mathbf{E}=(0,0,E_{z})$, it modulates the barrier height since $V_{\rm G}\rightarrow V_{\rm G}^{(0)}+V_{\rm G}^{(1)}\cos \left( \Omega t\right) $ [$V_{\rm G}^{(0)}$ is the steady state gate voltage, $V_{\rm G}^{(1)}$ is the a.c. field induced addition, see sketch in Fig.~\ref{fig:GQW2}(b)]. Then the a.c. field induced correction to the d.c. tunneling current is $j_{1}=2e \int d\varepsilon \chi _{\varepsilon }  \left\vert \delta t_{\varepsilon,\Omega }^{\left( 1\right) }\right\vert ^{2}\left( 2n_{\varepsilon }-n_{\varepsilon +\Omega -eV}-n_{\varepsilon -\Omega -eV}\right) $, where the transmission amplitude $t_{\varepsilon ,\Omega }^{\left( 1\right) }$ is
obtained from corresponding non-stationary boundary conditions at $x=0$ and $%
x=D$. Physically, the directional photoelectric effect (DPE) comes from an
ingenuous influence of the external electromagnetic field to the
electron-hole phase correlations during the chiral tunneling. Technically,
modulation of the barrier height by the a.c. field shifts positions of the sharp
peaks in the energy-dependent barrier transparency $T\left( \varepsilon \pm
\Omega \right) $. Besides, it also modifies the overall angular distribution of the
electric current, so the electron-hole conversions occur with an additional
phase shift. Numerical results for both the cases are presented in Fig.~\ref%
{fig:GQW2}(c,d). Corresponding plots for the steady state differential
conductance $\sigma _{0}(V_{\rm sd})$ and for the photon-assisted chiral tunneling conductance $\sigma
_{1}(V_{\rm sd})=\partial j_{1}/\partial V_{\rm sd}=\overline{\sigma}^t - \sigma_0$ both indicate the angular
redistribution of the photon-assisted chiral tunneling current across the graphene barrier. The steady
state conductance curve $\sigma _{0}$ in Fig.~\ref{fig:GQW2}(c) corresponds
to $U_{0}=2$~meV while curve $\sigma _{1}$ is computed for $V_{\rm G}=1$ V and $%
\Omega =1$~THz. The DPE is well illustrated by the sharp scattering resonance taking place in $\sigma _{1}(V_{\rm sd})$ [see the crisp peak at the incidence angle $\phi =\pi /16$ and at the bias voltage $V_{\rm sd}=U_{0}=0.5$
in Fig.~\ref{fig:GQW2}(c)]. When the a.c. field is off, the steady state
tunneling at $V_{0}=0.5$ in the straightforward direction is suppressed [see
the corresponding curve $\sigma _{0}\left( V_{\rm sd}\right) $ for $\phi =0$].
However, if one applies the a.c. field with frequency $\Omega $ and $\mathbf{%
E}=(E_{x},0,0)$, it opens tunneling channels in the straightforward
direction $\phi =0$ as is evident from curve $\sigma _{1}$ in Figs.~\ref%
{fig:GQW2}(c). In Fig.~\ref{fig:GQW2}(d) we compare two time-averaged
conductance curves $\sigma _{1}\left( V_{\rm sd}\right) $ under influence of
the a.c. field with two different polarizations along the $\hat{x}$ (curve
X) and $\hat{z}$ (curve Z) axes correspondingly. In either case the $\sigma
_{1}\left( V_{\rm sd}\right) $ curves show remarkable sharp peaks, which
position however changes versus the field polarization. 
\begin{figure}[tbp]
\includegraphics{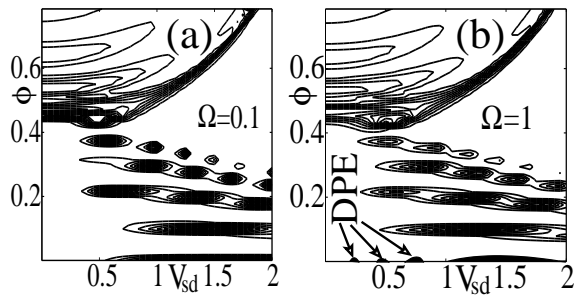}
\caption{Contour plots of the time-averaged differential conductance $\overline{\sigma (t)}^{t}$ (in units of $\tilde{\protect\sigma}_{0}$) of a bilayer graphene junction versus $V_{\rm sd}$ and $\phi$ at two different a.c. field frequencies (a)~$\Omega /U_{0}\simeq 0.1$ and (b)~$\Omega /U_{0}\simeq 1$.}
\label{fig:E_vs_phi_cont}
\end{figure}
Although the above results are illustrative, they focus solely on the limit of a weak
electromagnetic field $U_{1}<<U_{0}$. Influence of an external
electromagnetic field of arbitrary amplitude on the a.c. transport
properties of a bilayer graphene junction had been studied in this work
using a numeric approach. We solved the non-stationary boundary conditions
using the trial function (\ref{trial_f}) where we take $\mathcal{E}%
\rightarrow \mathcal{E}_{0}+\mathcal{E}_{1}\cos \left( \Omega t\right) $
with an arbitrary ratio $\mathcal{E}_{1}/\mathcal{E}_{0}$. We emphasize again that a mere multiphoton approximation like used in Ref.~\onlinecite{My-PRB} is not working in this case. The graphene barrier transparency now is not assumed to be small, therefore the electron wavefunction cannot be simply factorized as $\hat \Psi (x,t) \rightarrow \hat \Psi (x) \sum_k J_k \left(\alpha \right)\exp{(i n \Omega t)} $. Therefore we use a straightforward numeric solution of the non-stationary boundary conditions for $\hat \Psi (x,t)$ and compute the time-dependent transmission probability $T\left( t\right) $ directly from that solution. Then we apply a fast Fourier transform algorithm for computing of $T\left( \omega \right) $ numerically versus the external field frequency $\Omega $ and the a.c. barrier amplitude $U_{1}$. The obtained results for the differential conductance under influence of a strong electromagnetic field with $\Omega/U_{0}=0.75$ are presented in Fig.~\ref{fig:E_vs_W}. In Fig.~\ref{fig:E_vs_W}(a) we show the time-averaged conductance $\overline{\sigma }=\overline{\sigma \left( t\right) }^{t}$ of the bilayer graphene junction for the normal electron incidence $\phi =0$ and for
different a.c. field amplitudes $U_{1}=0.01$ (curve A), $U_{1}=0.1$ (curve
B), $U_{1}=0.4$ (curve C), and $U_{1}=1.3$ (curve D). One can see that if the external field amplitude $U_{1}$ is lower than the graphene barrier height $U_{1}<U_{0}$ (which corresponds to curves A-C) the
junction$^{\prime}$s conductance has a threshold character versus the bias voltage $%
V_{\rm sd}$. If, however, $U_{1}>U_{0}$, a finite transparency takes place even at $V_{\rm sd}<U_{0}$,
which corresponds to curve D. From the three-dimensional plot $T_{0}\left(
\varepsilon ,\Omega \right) $ shown in Fig.~\ref{fig:E_vs_W}(c) one can see that a visible
transparency is achieved at frequencies $\Omega /U_{0}\approx 1$, which is
well consistent with the semi-qualitative consideration above. A more
accurate estimation of the DPE magnitude follows from Fig.~\ref{fig:E_vs_W}(d) where we
plot $\overline{\sigma }\left( V_{\rm sd}\right) $ for three different frequencies $\Omega
/U_{0}=1$ (curve 1), $\Omega /U_{0}=0.1$ (curve 2), and $\Omega /U_{0}=2$
(curve 3). The peaks of finite $\overline{\sigma }$ in curves 1-3 at $E_k<U_0$ are present because the electron incidence angle $\phi $ is finite though small ($\phi = \pi/20$). The peak increase of the junction$^{\prime }$s conductance $\sigma (V_{\rm sd}^{(k)})$ is achieved at selected bias voltage values $V_{\rm sd}^{(k)}<U_{0}/e$ and $\Omega /U_{0} \approx 1$, which corresponds to curve 1. One can see that the directional photo-electric effect  increases the junction conductance $\sigma (V_{\rm sd}^{(k)})$  at $\Omega /U_{0} \approx 1$ by a few orders of magnitude as compared to its steady state value at the same $V_{\rm sd}^{(k)}$. Relevant increase of  the conductance is however less significant at other a.c. field frequencies, i.e., $\Omega /U_{0}=0.1$ (curve 2), and $\Omega /U_{0}=2$ (curve 3). In Figs.~\ref{fig:E_vs_phi_cont}(a,b)  we show contour plots of the time-averaged conductance $\overline{\sigma (t)}^{t}$ of a bilayer graphene junction versus the electron incidence angle $\phi$ and the source-drain bias voltage $V_{\rm sd}$ for two different a.c. field frequencies (a) $\Omega /U_{0}\simeq 0.1$ and (b) $\Omega /U_{0}\simeq 1$.  The conductance diagrams in either case  have pretty spectacular complex structure where the dark spots correspond to $\overline{\protect\sigma}^t \simeq \tilde{\protect\sigma}_{0}$. When the external field frequency $\Omega $ is low [i.e., $\Omega/U_0 = 0.1$ as shown in Fig.~\ref{fig:E_vs_phi_cont}(a)], the tunneling for the incident electron energies $E/U_0 < 0.55$ is fully blocked.  However, when the field frequency becomes higher, i.e., $\Omega /U_{0} = 1$ as indicated in Fig.~\ref{fig:E_vs_phi_cont}(b), one may notice a series of sharp dark spots at discrete energies $E_k$ below the barrier ($E_{k}<U_0$) pronounced at the normal incidence angle $\phi=0$. Those dark spots constitute the directional photoelectric effect discussed above and indicated as DPE in Fig.~\ref{fig:E_vs_phi_cont}(b).

Intrinsic noise in the bilayer graphene junction originates as follows. The
thermal noise comes from the phonons emitted in the electron-phonon
collisions. Matrix element of the electron-phonon collisions according to
Ref. \cite{Ando, McEuen} is $M_{p p^{\prime}}\propto \left\langle
p\left\vert M\left( x\right) \right\vert p^{\prime}\right\rangle \cos \left(
\phi _{p p^{\prime}}\right) $ where $\phi _{p p^{\prime}}$ is the angle
between initial and final states. The phase factor $\cos \phi _{p
p^{\prime}}$ plays quite a different role in the bilayer graphene 
compared to the single layer graphene\cite{Ando} where it is rather 
$\cos \left( \phi _{p p^{\prime}}/2\right) $ instead. In the latter case,
the factor ensures suppression of the electron-phonon and electron-impurity
collisions and the transport of the change carriers remains ballistic up to
room temperatures. In contrast, thermal noise in the bilayer graphene
devices is rather high at room temperatures. Another intrinsic noise
(Poisson noise) arises due to the "Zitterbewegung" effect, which is linked
to a jittering motion of the change carriers when electrons are randomly
converted to holes forth and back. That produces noise even in zero
temperature limit. The noise is characterized by the Fano factor $F=\sum_{n}T_{n}%
\left(1-T_{n}\right) /\sum_{n}T_{n}$, where $T_{n}$ is the tunneling
probability in the $n$-th channel and the summation is performed over all
the conducting channels (in our setup this means just integration over $\phi$%
). From the plot $F (V_{\rm sd})$ shown in Fig. \ref{fig:GQW2}(d) for $D=15$ (in
units of $h/\sqrt{2mU_0}$) one infers that the Poisson noise becomes extremally
low at $V_{\rm sd}\geq U_0$.

\section{Conclusions}
In conclusion we computed the electric current across the bilayer graphene
junction in conditions when an external electromagnetic field is
applied. We have found that the threshold absorption of the external
electromagnetic field strongly depends on the a.c. field frequency and amplitude.
The electromagnetic field induces an ideal transparency of the graphene barrier in the longitudinal
direction, which had been fully suppressed when the a.c. field was off. That directional photoelectric effect originates from an angular redistribution of the whole transparency diagram since the sidebands at finite angles are redirected to the normal incidence. An experimental observation of such a spectacular directional optoelectric
phenomena would provide a strong evidence for existence of the massive chiral fermions in
the bilayer graphene. We emphasize that the threshold absorption emerges purely from a quantum mechanical phase shift, and not from an inelastic excitation by the a.c. field. That means no heating is involved during the
absorption. The a.c. current induced by the electromagnetic field across the graphene junction
has a sharp angular dependence, which potentially can be exploited in sensor
nanodevices of the external electromagnetic field. The directional
photoelectric effect in the double layer graphene junctions is a unique
phenomenon which exists in that system and had not been noticed in other
systems, like junctions composed of single layer graphene or of normal metals.
Most intriguing feature is the switch between zero and finite
conductance occurring without energy absorption. The phenomena considered
above have a great potential for various nanoelectronic applications.

I wish to thank V. Chandrasekhar and P. Barbara for fruitful discussions.
This work had been supported by the AFOSR grant FA9550-06-1-0366.

\end{document}